\documentclass[preprints,article,accept,moreauthors,pdftex]{mdpi}

\firstpage{1} 
\pubvolume{xx}
\issuenum{1}
\articlenumber{5}
\pubyear{2020}
\copyrightyear{2020}
\history{Received: date; Accepted: date; Published: date}
\Title{
A Thermodynamic Approach towards the Question “What is Cellular Life?”
}

\Author{Yasuji Sawada $^{1}$*, Yasukazu Daigaku $^{2,3}$ and Kenji Toma $^{1,2,4}$}

\AuthorNames{Yasuji Sawada, Yasukazu Daigaku and Kenji Toma}

\address{%
$^{1}$ \quad Division for Interdisciplinary Advanced Research and Education, Tohoku University, Sendai 980-8578, Japan\\
  $^{2}$ \quad Frontier Research Institute for Interdisciplinary Sciences, Tohoku University, Sendai 980-8578, Japan\\
  $^{3}$ \quad Graduate School of Life Sciences, Tohoku University, Sendai, Japan\\
  $^{4}$ \quad Astronomical Institute, Tohoku University, Sendai, 980-8578, Japan
}

\corres{Correspondence: yasuji.sawada.a4@tohoku.ac.jp}
\abstract{
The question “What is life?” has been asked and studied by the researchers of various fields. Nevertheless, no global theory which unified various aspects of life has been proposed so far.  Considering that the physical principle for the theory of birth should be the one known for the inanimate world, and that the life processes are irreversibly selective, we showed by a deductive inference that the maximum entropy production principle plays an essential role for the birth and the evolution of life in a fertile environment. In order to explain the survival strategy of life in a barren period of environment, we also proposed that life had simultaneously developed a reversible on and off switching mechanism of the chemical reactions by the dynamics of equilibrium thermodynamics. Thus, the birth and evolution of life have been achieved by the cooperation between the driving force due to the non-equilibrium thermodynamics and the protective force due to the equilibrium thermodynamics in the alternating environmental conditions. 
}

\keyword{
birth and evolution of life; non-equilibrium thermodynamics; maximum entropy production principle; self-replication; multi-cellular life
}

\begin{document}
%%%%%%%%%%%%%%%%%%%%%%%%%%%%%%%%%%%%%%%%%%

%%%%%%%%%%%%%%%%%%%%%%%%%%%%%%%%%%%%%%%%%%
\section{Introduction}
\label{sec1}
\begin{figure}
\centering
\includegraphics[width=15cm]{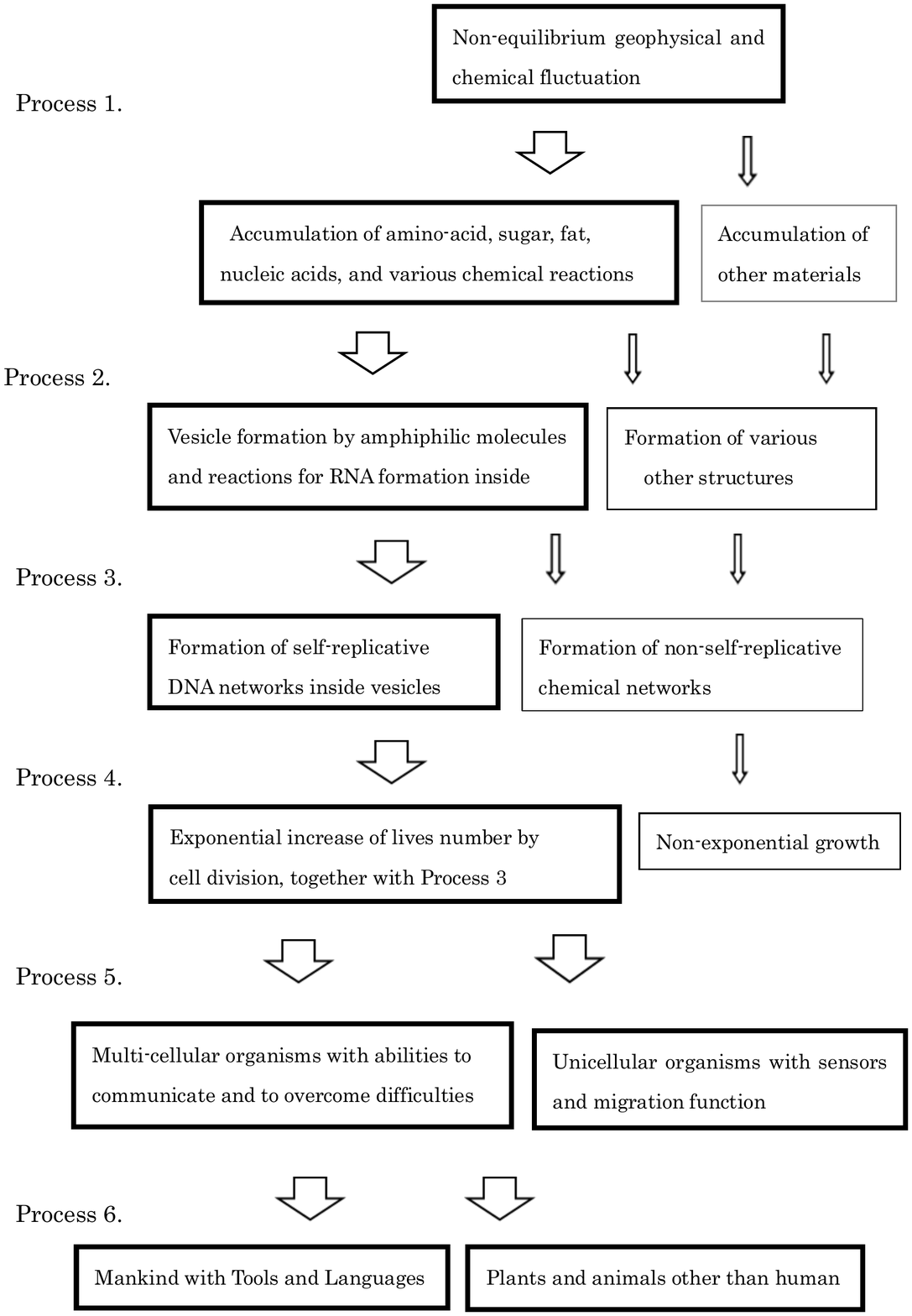}
\caption{
Breakthrough processes of the birth and evolution of life. Heavy frames and thick arrows are the main line of birth and evolution of lives. The figure is by no means complete, with many other missing important steps of evolution such as the transition from asexual to sexual regeneration in this figure.
}
\label{fig1}
\end{figure}

The structures of biological organisms are miraculously complex and the functions are extremely multi-fold far beyond imagination. Even the global features such as metabolism, self-replication, evolution or common cell structures make us wonder how they were created in the history of nature. Figure~\ref{fig1} shows the processes generally accepted as important for the birth and evolution of life.

Accidental accumulation of amino-acid, sugar, fat, nucleic acids in the varieties of planetary and geophysical phenomena such as meteorite bombardment, volcanic activities, temperature fluctuation occurred on the surface of the earth, prepared a stage for the birth of life [Process 1]. Among the quasi-stable amphiphilic vesicles produced in different environments, some happened to contain density of useful chemicals high enough to set the cells in the thermodynamic state far from equilibrium and various chemical reaction modes were excited inside, including formation of RNA world [Process 2]. The vesicles finally became biological cell membrane, which had provided a stable dense chemical reaction tanks. The well-developed cell structure started developing self-replicating double-stranded DNA with catalysis by proteins or other chemical compounds [Process 3]. The exponential growth in number of life by the self-replication made visible the very small probability of natural mutation and also made possible the “Preservation of Favoured Races in the Struggle for Life” by Darwin \cite{darwin1859} [Process 4].  Frequent cell division gave a choice of life style whether as independent single cells or as cooperative group cells, depending on the environmental condition [Process 5].  Invention of tools and acquisition of languages of mankind [Process 6].

Since the discovery of the molecular structure of DNA by Watson and Click \cite{watson53}, enormous progress of experimental studies of living organism has been made based on DNA central dogma in molecular biology. In contrast, theoretical biology in questioning how the life was born and how it has evolved is less advanced. Physicists had been eager to find some fundamental concepts in this charmingly complex world \cite{oparin38,schrodinger44,prigogine68,glansdorff71,kondepudi14,eigen71,eigen13,dyson85,kauffman95,walker17}. Schrödinger \cite{schrodinger44} discussed the “negative entropy” of the life which was a new concept at that time, but it was not explained where living organisms come from and how they are generated, in spite of constructive model proposed by experimental researchers such as RNA world \cite{gilbert86,szostak16}. Prigogine \cite{prigogine68} derived formalism of entropy production and its fluctuation and discussed stability of the local steady non-equilibrium systems. Eigen \cite{eigen71,eigen13} who kept through his life asking the physical meaning of Darwin’s natural selection \cite{darwin1859} and tried to answer the question “Who does the selection?” \cite{eigen13} We focus on the physical principle by which nature selected a path shown by heavy frames and thick arrows among others in Figure~\ref{fig1}.

In this paper, we start by confirming that three basic points are indispensable for the comprehensive understanding of the birth and evolution of life [from Process 1 to 5] shown in Figure~\ref{fig1}. First, the birth process must be based on a principle governing inanimate world, because life was basically born from inanimate material world. Second, an essential feature of life is a metabolic activity associated with an open system to the environment. These two points strongly imply that the life phenomena should be based on the thermodynamics of open non-equilibrium systems. Third, if so, we must ask if self-replication, the most fundamental mechanism of the life in addition to the metabolic activity, can be explained by the thermodynamics of an open system far from equilibrium. We demonstrate for the first time in this paper that the maximum entropy production principle \cite{ziegler63,sawada81}, which has been shown to work as the selection principle in the inanimate phenomena far from equilibrium, plays an important role for answering the third point as well as the evolution of living organism and the associated activities.

In addition, we discuss what happened to the lives in a barren environment when they failed to satisfy the thermodynamic condition far from equilibrium [Process 5]. The life had simultaneously developed a reversible on/off switching mechanism of chemical reactions to survive in the barren environment.

The structure of this paper is composed of an introductory part of the maximum entropy production in Section \ref{sec2}, and the deductive results for the birth and evolution of life from the principle of maximum entropy production in Section \ref{sec3}. In Section \ref{sec4}, we discuss thermodynamics of multi-cellular organism together with the altruistic behavior developed for a barren condition. In Section \ref{sec5}, we discuss the remaining problems and conclude.

%%%%%%%%%%%%%%%%%%%%%%%
\section{Thermodynamic approach to a complex open system far from equilibrium}
\label{sec2}

\subsection{Maximum entropy production principle}
This principle has been proposed independently \cite{ziegler63,sawada81} for a local system connected weakly to the reservoirs which are far from equilibrium from each other. The local system plays the role of transporting entropies between the reservoirs. The loss of entropy by this transport, which is equivalent to the “negative entropy” of Schrödinger, is compensated by the entropy production within the local system to keep a steady state. The entropy of the reservoirs increases by the amount of the transported entropy through the local system which is equal to the entropy production of the local system. Thus, the thermodynamics of the reservoirs directly influence the behavior of the local system. The principle states that nature chooses the mode for the local system whose entropy production is highest among possible modes (see Appendix~\ref{app1} for an intuitive derivation of the principle). This principle works equally for the local open systems, whether they are inanimate, prebiotic, or animate, if the thermodynamic condition is far from equilibrium. That is why this principle is useful for studying the origin and evolution of life. This principle may serve as the physical basis of “Survival of the Fittest” of C. Darwin \cite{darwin1859} when the environment satisfies a thermodynamic condition far from equilibrium.

\subsection{Past researches in inanimate systems based on the maximum entropy production principle}

The maximum entropy production principle has been applied for the self-organized dynamical structures and the corresponding functions of open systems far from equilibrium \cite{kleidon04,martyushev06}, as the natural extension of the second law of thermodynamics. In the studies of pattern formation of a chemical reaction \cite{shimizu83} and electrical convection patterning \cite{suzuki83}, computer simulation showed that the pattern modes drifted from given initial conditions towards the mode of largest entropy production among the possible modes under the given thermodynamic potential. A vast of researches on the climate and turbulence in relation to the maximum entropy production were reported \cite{paltridge79,paltridge81,lorenz01,shimokawa02,ozawa01,malkus54}. Thus, the total system consisting of the local open system together with the reservoirs were observed to follow the maximum entropy production principle. This principle and its use have been discussed by researchers of the field \cite[e.g.][and the other papers in the special issue of Entropy ``What Is Maximum Entropy Production and How Should We Apply It?" edited by Kleidon, A. \& Dyke, J. (2009)]{dewar09}.

Recently, a theoretical research of the selection rule of open systems far from equilibrium was reported based on the stochastic least-action principle, which showed that in a multi-stable system, the steady state with the highest entropy production is favored \cite{endres17}. Among statistical theories of evolution, a model \cite{rivoire14} was recently proposed to give the relation between the maximum growth rate of individual number and statistical variations of the attribute, the offspring and the environmental factors. The maximum growth rate is certainly related to the maximum entropy production. A statistical approach may be powerful for discussing evolution once we obtain the transition probability.

The present status of the researches in the field of non-equilibrium physics of biology was recently reviewed in detail \cite{fang19}, which showed the non-equilibrium nature of each biological phenomena. It also suggested that the researches have been localized in each phenomena of biology, and have not yet approached to the present global question of the life.

%%%%%%%%%%%%%%%%%%%%%%5
\section{Deductive outcomes in prebiotic and biological organization from maximum entropy production principle}
\label{sec3}

\subsection{Entropy production and chemical structures in the open chemical reaction systems far from equilibrium}

\begin{figure}[H]
\centering
\includegraphics[width=15cm]{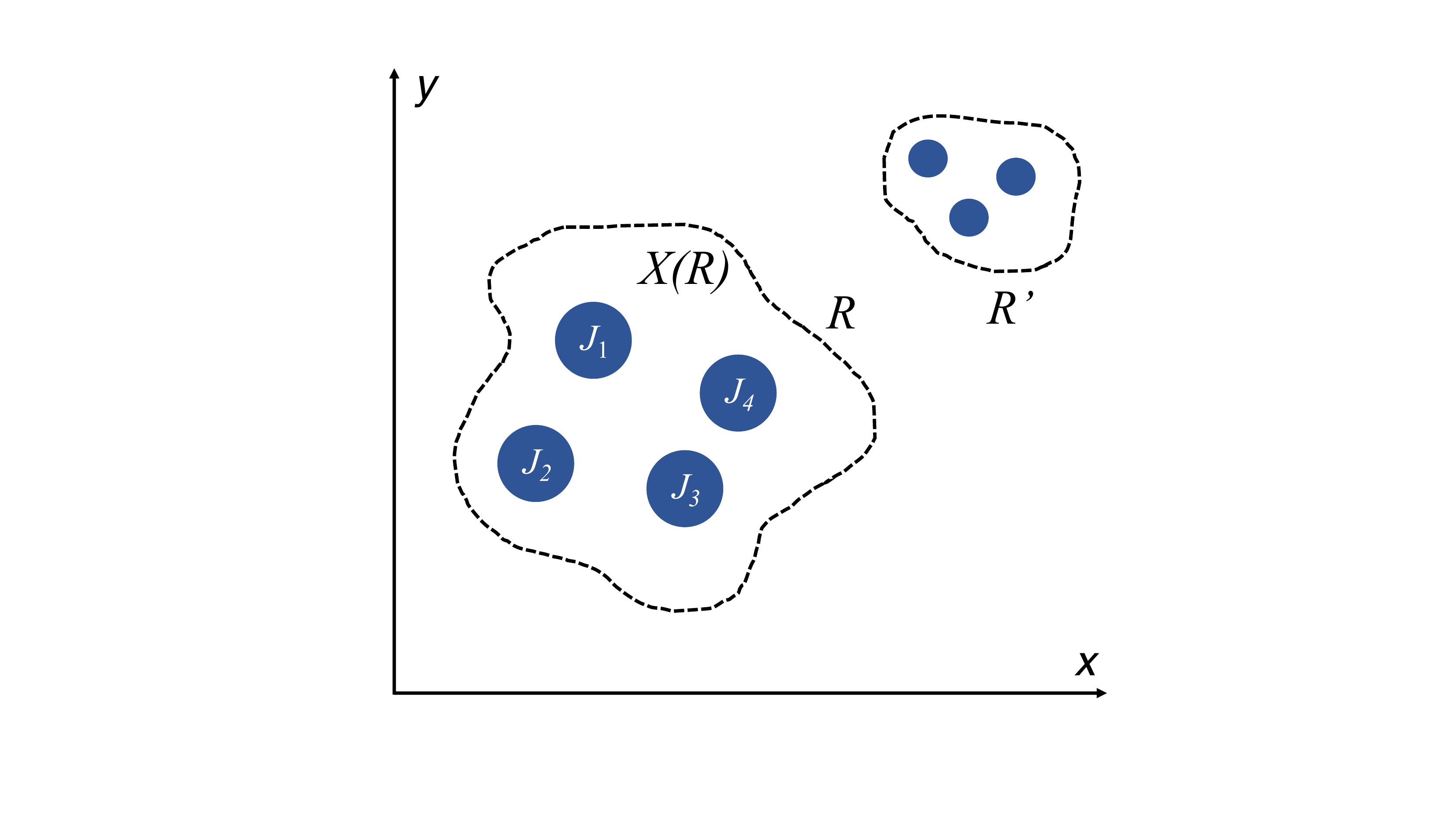}
\caption{
Schematic drawing for a thermodynamic system at niches $R, R', ...$, far from equilibrium, consisting of a reservoir which is characterized by a thermodynamic force $X(R)$ and some small open organisms with thermodynamic currents $J_i$. In a prebiotic or a biological organism, further non-equilibrium thermodynamic relation exists within each organism between the chemical components (see texts in Section~\ref{subsec3-3}).
}
\label{fig3}
\end{figure}

The principle of maximum entropy production is concerned with the amount of entropy production of a local system weakly connected with one or more large reservoirs, with conditions that the local system is far from the equilibrium with the reservoirs. The local system, in general, may be separated from the reservoirs in real space, momentum space or the chemical component space. The local system of our interest is some assemblies of chemical components, which is weakly coupled with the environment (see Figure~\ref{fig3}). Simultaneously, each assembly is composed of some chemical components whose numbers changes by reactions in the direction to increase entropy. An example of the separation in chemical component space is one box of the same temperature, but with two different kinds of molecules $A_1$ and $A_2$ with different chemical potentials $\mu_1$ and $\mu_2$, interchangeable irreversibly from one to the other by reaction. When a chemical reaction from $A_1$ to $A_2$ or vice versa is allowed, the entropy production due to the reaction during a time interval $dt$ is given by
\begin{equation}
\sigma dt = -\mu_1 dN_1 - \mu_2 dN_2,
\end{equation}
where $dN_1$ and $dN_2$ are the number changes of each molecule by the reaction during $dt$.

More generally, the entropy production $\sigma(R,t)$ of an ensemble of organisms in the area $R$ which is much larger than the body size of the organism (see Figure~\ref{fig3}), may be written at time $t$ as,
\begin{equation}
  \sigma(R,t) = \sum_i \sum_k J_{i,k}(R,t) \cdot X_{i,k}(R,t)
\label{eq:sigma}
\end{equation}
where $J_{i,k}$ and $X_{i,k} = A_{i,k}/T$ are the reaction rate and the generalized thermodynamic force for the $k$-th chemical reaction of i-th organism, respectively \cite{glansdorff71,kondepudi14}. And, the chemical affinity $A_{i,k}$ is written as,
\begin{equation}
A_{i,k} = -\sum_j v_{i,k,j} \mu_{i,k,j},
\end{equation}
where $v_{i,k,j}$ is the $j$-th component of $k$-th reaction of $i$-th organism. $\mu_{i,k}$ is the chemical potential of $k$-th reaction of the $i$-th organism.

We note here that the thermodynamic current realized at a condition far from equilibrium generally forms a structure in the relevant space. For a non-equilibrium system whose thermodynamic force is separately given in a real space, a spatial current structure such as convection is observed. While, for a non-equilibrium system whose thermodynamic force is separately given in a chemical component space, an associated structure of reaction flows will be observed in chemical components space. The self-replication chemical reaction structure in the problem of birth and evolution of life is also an example of the structure in a chemical component space generally associated for the maximum entropy production principle in a non-equilibrium system far from equilibrium as we show in a later section.

\subsection{Thermodynamic deductive scenarios for the birth and evolution of life}
In a simple thermodynamic system where the dynamical equation can be written down, such as the case of a convection system, or the case of model chemical reaction system such as Brusselator \cite{prigogine68,glansdorff71}, the thermodynamic force $X_k$ and the thermodynamic current $J_k$ can be specified. On the other hand, chemical reaction systems related to the birth and evolution of life are so complex that it is nearly impossible to specify all the dynamics of reactions. Since any bottom up theoretical treatment may be effective only for individual phenomena, we present here a top down thermodynamic approach for the comprehensive understanding of life. As described in Section~\ref{sec1} and \ref{sec2}, the candidate of the top down principle is the maximum entropy production principle. Thus we examine in this section whether various processes of life can be deduced from this principle, and compare with the processes observed (see Figure~\ref{fig1}) for the birth and evolution of life.

By limiting the discussion here for only one kind of organisms in a local area, and by summing Equation (\ref{eq:sigma}) over $i$, we get,
\begin{equation}
  \sigma(R,t) = N(R,t)\langle J(R,t)\cdot X(R,t)\rangle
\label{eq:sigma2}
\end{equation}
where $N$ is the number of organisms, and $\langle J \cdot X\rangle$ is the inner product of vectors $J$ and $X$ in $k$-space. Since thermodynamics of each area is independent from each other when each area is not interacting, one can discuss the entropy production of an area and the principle of maximum entropy production requires to maximize $\sigma$ of each local area. From Equation (\ref{eq:sigma2}) the change of the entropy production $\sigma(R,t)$ has three terms,
\begin{equation}
  \Delta\sigma(R,t) = \Delta N \langle J(R,t) \cdot X(R,t) \rangle +
  N(R,t) \langle \Delta J(R,t) \cdot X(R,t) \rangle +
  N(R,t) \langle J(R,t) \cdot \Delta X(R,t) \rangle
\label{eq:sigma3}
\end{equation}
Equation (\ref{eq:sigma3}) implies that maximum entropy production principle is decomposed into three scenarios.

[$\Delta N$–scenario.] The first term of Equation (\ref{eq:sigma3}) is the contribution to the increase of the entropy production by increasing number $N$. The number of the subsystems may increase even in inanimate systems such as vortices in fluid system, although the mechanism of the birth of a new eddy is independent of each other, while the specialty of the $\Delta N$–scenario of the prebiotic and biological systems is the achievement of self-replication, and therefore the born subsystems are identical to each other. The chemical reactions for the self-replication is discussed in the following section.

[$\Delta J$-scenario.] The second term of Equation (\ref{eq:sigma3}) is the contribution to the increase of the entropy production by increasing thermodynamic current $J$. The chemical reaction rates were improved by the transition of the modes to higher entropy production shown in Figure \ref{fig2} introducing various kinds of enzymes and new reactions before or after the completion of the self-replication cycle. Geometry, membrane and size of the cells were adjusted for the highest efficiency of the metabolism after the self-replication cycle was completed.

[$\Delta X$–scenario.] The third term of Equation (\ref{eq:sigma3}) corresponds to improving the mismatch of the position $R$ of the cell and the more fertile environment,
\begin{equation}
\Delta X(R,t) = \Delta R \cdot \nabla_R X(R,t).
\end{equation}
The cell may move to a better environment to increase entropy production if the cell is equipped sensors to sense $\nabla_R X(R,t)$ and mobile function to realize $\Delta R$. For this strategy to function for a cell, the sensor and locomotive mechanism must be developed.

Now, let us try to compare these deductive possibilities with the processes shown in Figure \ref{fig1}. We realize that the Process 1 and 2 correspond to $\Delta J$-scenarios, and that Process 3 and a part of Process 4 correspond to $\Delta N$–scenarios. Process 5 may correspond to a $\Delta X$–scenario. The three scenarios were obviously not planned by the biological system. It is only the result of the selection among the products of various mutation to satisfy the maximum entropy production principle. Process 6 is obviously beyond the scope of the present paper.

\subsection{Self-replication in chemical systems and the maximum entropy production principle}
\label{subsec3-3}
In the previous section we deductively show that the main processes of birth and evolution of life is consistent with the maximum entropy production principle. This deductive approach is to show a comprehensive understanding of various phenomena observed in the process of biological life. To obtain the concrete mechanism of each process, a dynamical model consistent with the principle is useful.

We show hereafter the self-replication mechanism, the most important process for the birth of life, is consistent with the maximum entropy production principle. One cannot expect a complex molecular structure to make a direct copy of itself, because a molecule of one kind avoid to be near the same molecules. However, one can expect to make a copy, first by making a negative copy which is complemental to the original, and second, by making a negative copy of the negative. If this cycle works well ($\Delta J$-scenario), the original complex molecular structure will be doubled, and the continuation of this process will lead to exponential increase of $\Delta N$ ($\Delta N$-scenario) by creating a fastest reaction mechanism of this self-reproduction process shown in Appendix~\ref{app2} and in Figure \ref{fig4}. Chemical reaction current in a chemical component space of the self-replication dynamics is shown in Appendix~\ref{app2} under a thermodynamic condition far from equilibrium. $X$ and $X^*$ are the single strand gene polymer and its counterpart. $XX^*$ and $X^* X$ are equivalent double strand gene polymers. $C$ is law molecules, and $E$, $E'$, and $E''$ are enzymes.

We notice a surprising similarity between the self-replication reaction currents in the chemical component space (Figure~\ref{fig4}) and the convection currents of fluids in the spatial coordinate (Figure~\ref{fig5}). They are both the results of the maximum entropy production of each system at a condition far from equilibrium. The self-replication phenomena might as well be called “chemical convection”. However there is an important difference between this chemical convection and fluid convection. In the former one cycle produces double entropy by doubling the number of $XX^*$, as each cycle is composed of double circulation lines(Figure~\ref{fig4}). In the latter (Figure~\ref{fig5}), the fluid motion constantly carry a constant amount of entropy from the hot bottom plate to the cold top plate.

\begin{figure}[H]
\centering
\includegraphics[width=11.5cm]{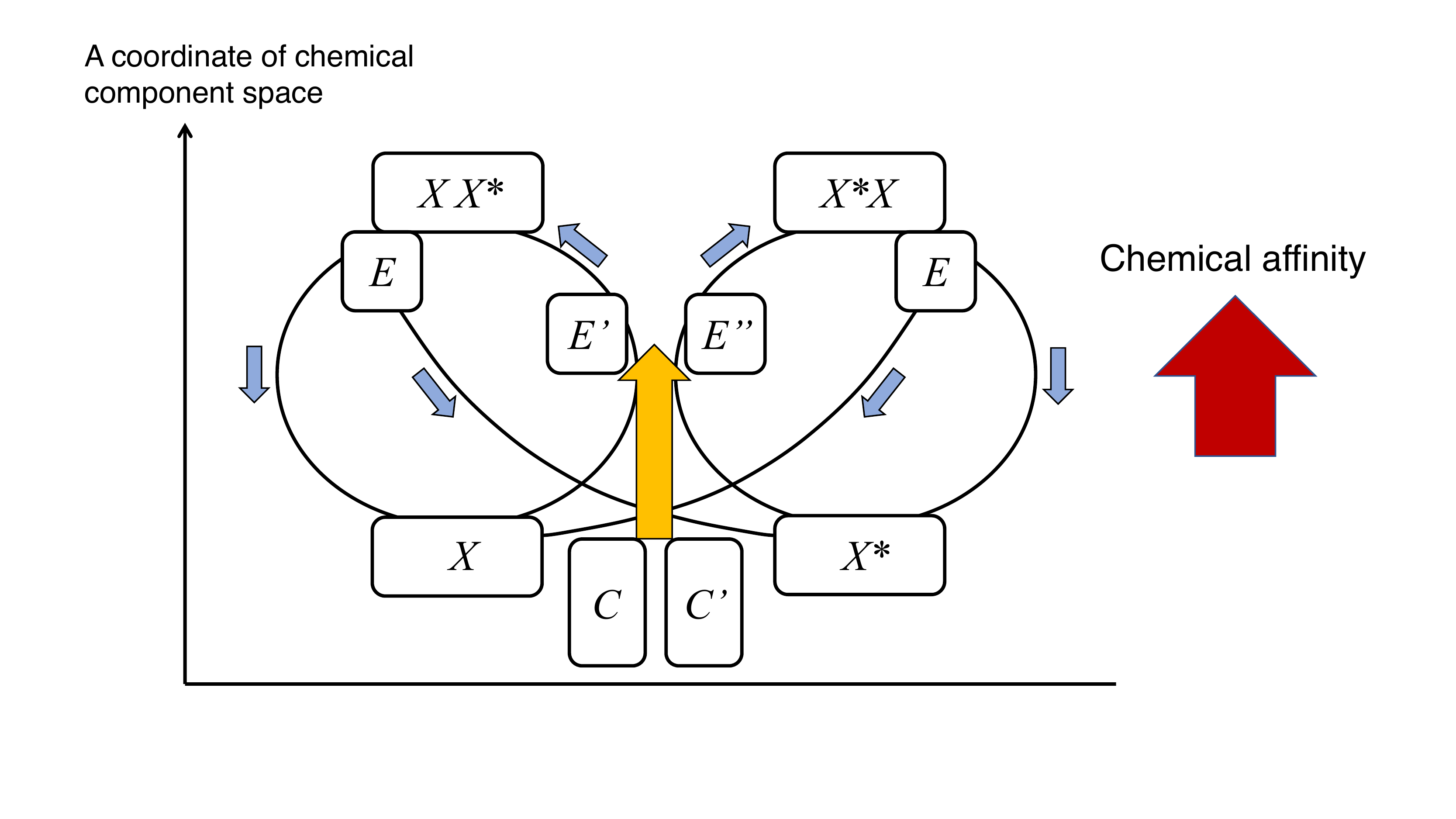}
\caption{
Chemical reaction current in a chemical component space of the self-replication dynamics shown in Appendix~\ref{app2} under a thermodynamic condition far from equilibrium. The chemical reactions are shown in Appendix~\ref{app2}. $X$ and $X^*$ are the single strand gene polymer and its counterpart. $XX^*$ and $X^* X$ are equivalent double strand gene polymers. $C$ is law molecules, and $E$,$E'$, and $E''$ are enzymes.
}
\label{fig4}
\end{figure}
\begin{figure}[H]
\centering
\includegraphics[width=11.5cm]{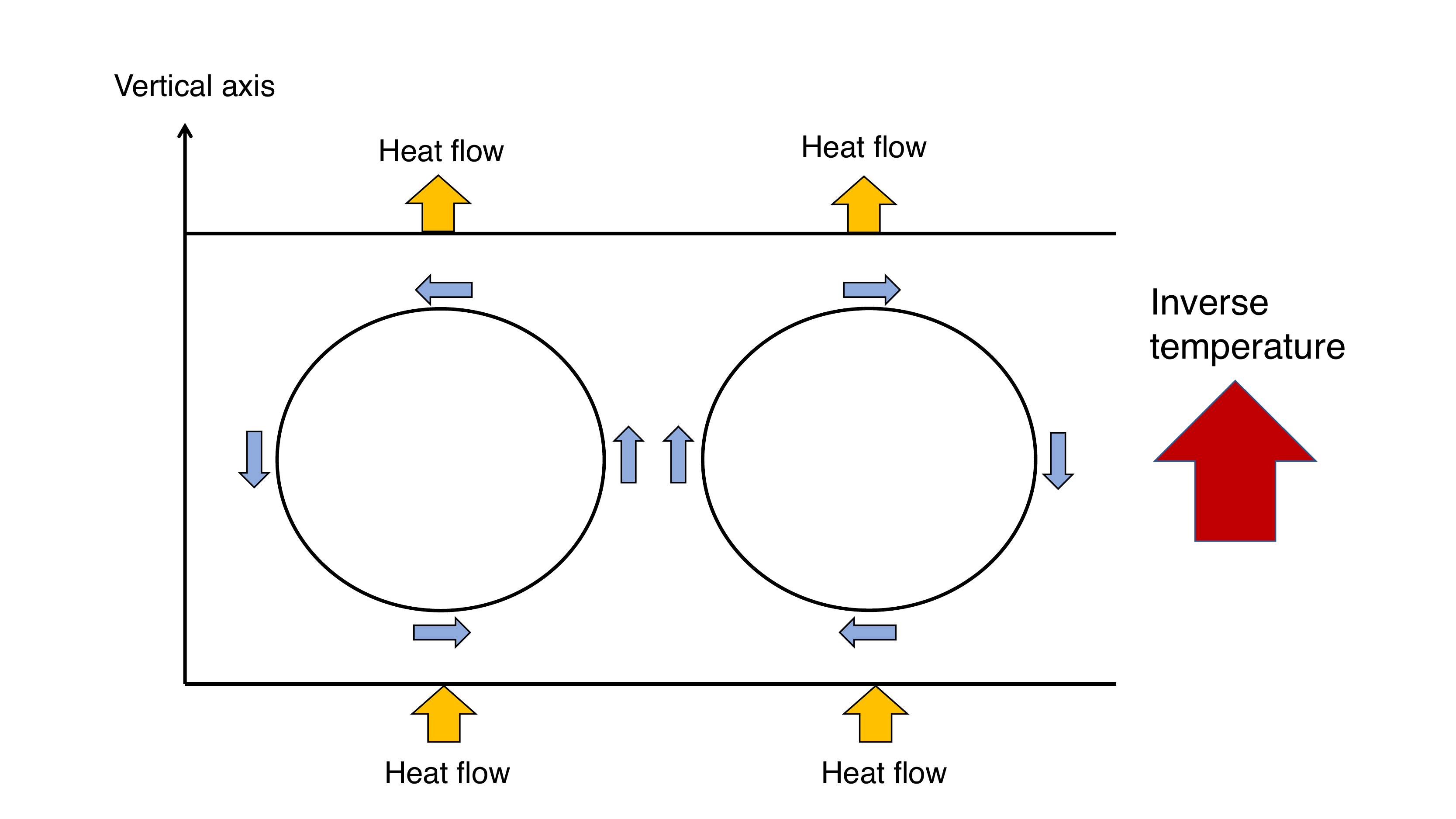}
\caption{
Convection current of a liquid layer in real space under a condition far from equilibrium.
}
\label{fig5}
\end{figure}

This chain reaction increases number of RNA or DNA exponentially with time (Appendix~\ref{app2}) and guarantees the fastest possible increase of the number $N$, and thereby, satisfies the maximum of entropy production. No other reactions can make the entropy production higher. Self-replication of RNA information was an indispensable step [Process 3 of Figure~\ref{fig1}] for the evolution of life, as the RNA is considered to be the first pre-biotic self-replicating molecules which carry information \cite{gilbert86,szostak16}.

\subsection{An example of possible scenarios for the birth of cellular life}
An important question is in what conditions the self-catalytic reaction can stably continue. First of all, there must be accumulation of high concentration of law materials for the self-catalytic reaction [Process 1 of Figure \ref{fig1}]. An open space may be difficult for this condition, as the law chemical molecules should be diluted in an open space by the water flow and winds. Pores in the rock might be better, which may be related to the DNA multiplication experiments by Braun et al. \cite{braun03}. The key point for the successful results of their experiments is due to the combinatorial usage of the principle of maximum entropy production principles, one in the chemical component space shown in Figure \ref{fig4} and the other in the real space shown in Figure \ref{fig5}. The temperature oscillation of the enzymes in the convective current (Figure \ref{fig5}) could be controlled to synchronize with the reaction times of melting and fusing of DNA molecules (Figure \ref{fig4}).

\begin{figure}[H]
\centering
\includegraphics[width=12cm]{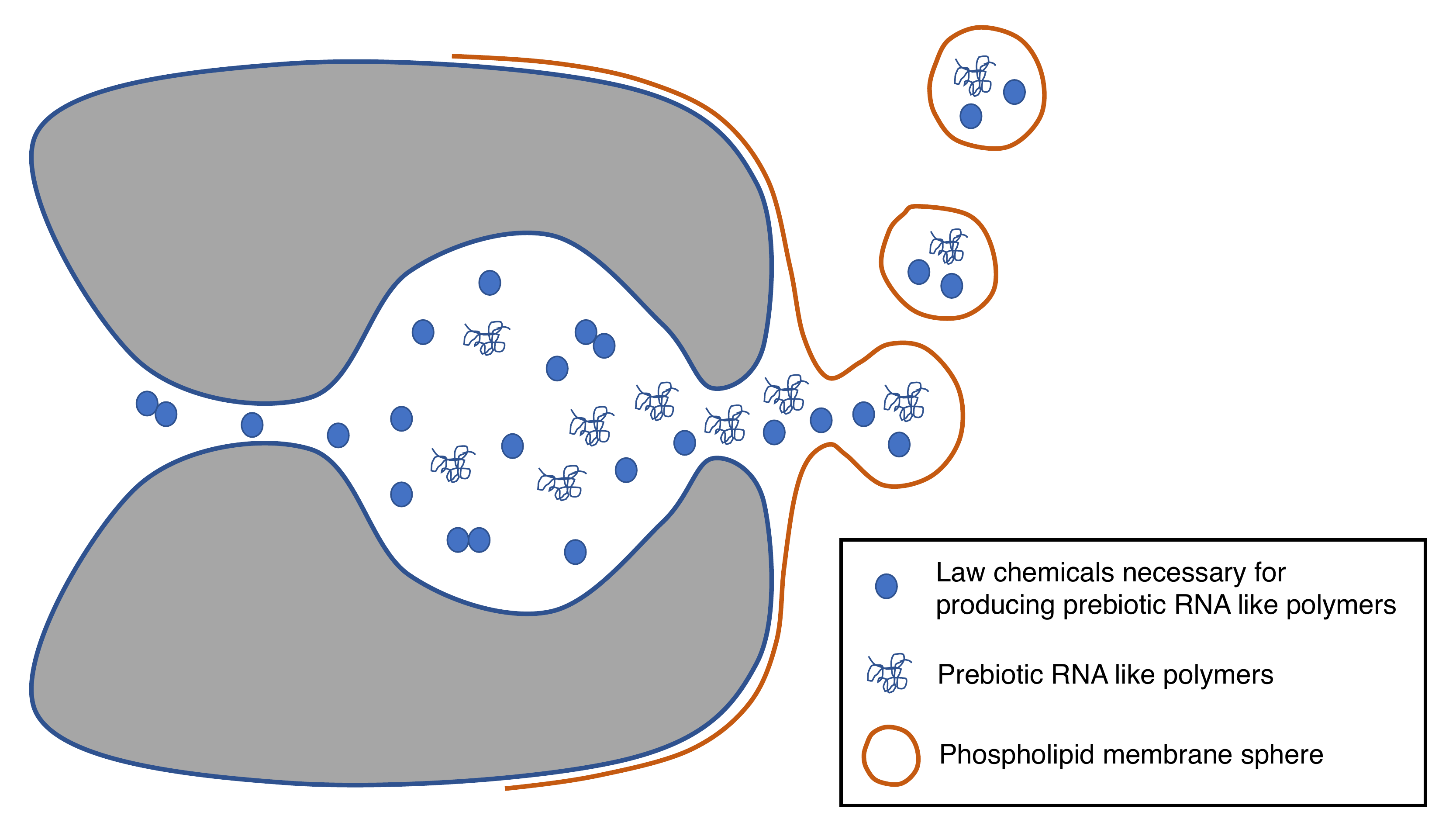}
\caption{
Schematic drawing of hypothetical formation of prebiotic self-replicating organism in a pore covered by phospholipid membrane. The produced self-replicated RNA like polymers flew out from the pore before the law chemical materials are depleted. If the exit of the window of the pore happened to be covered by phospholipid membrane, the self-replicated polymers could have been covered at the exit by the membrane balls like the soap bubbles [Process 2 and 3 of Figure \ref{fig1}]. The bubble membrane might have played the role of a vehicle to protect the self-replicated products until it finds a new environment for further evolution. And after a long time, this membrane finally might have grown into a biological cell membrane \cite{hardy15}, and Process 4 of Figure \ref{fig1} was realized.
}
\label{fig6}
\end{figure}

Secondly, the produced self-replicated polymers must flow out from the pore before the law chemical materials are depleted. For this purpose, small windows connected from the pore to the external world are necessary so that the exponentially increasing products can enjoy in a fresh and nourishing environment. Thirdly, if the exit of the window of the pore is covered by phospholipid membrane \cite{israelachvili11}, the self-replicated polymers may be covered at the exit by the membrane balls like the soap bubbles [Process 2 and 3 of Figure \ref{fig1}]. The membrane might have played the role of a vehicle for the self-replicated products until it finds a new environment for further evolution. And after a long time, this membrane finally might have grown into a biological cell membrane \cite{hardy15}, and Process 4 of Figure \ref{fig1} was realized. An example of this scenario is shown in Figure \ref{fig6}.

Great improvement of metabolism by using variety of enzymes were necessary [$\Delta J$-scenario.] for the Process 2, 3 and 4, and the additional functional structures such as lysosomes or mitochondria underlie the Process 5 of Figure \ref{fig1}. Paramecium \cite{alberts83} is known as a model unicellular life which has temperature sensor and cilia for swimming which may have been developed as one of [$\Delta X$–scenario] which is included in the Process 5 of Figure \ref{fig1}. The comparison shows that the most of the possible scenarios deductively driven from the maximum entropy production principle are found in the main processes of life at least of the unicellular organism.

%%%%%%%%%%%%%%%%
\section{Thermodynamic stability of the living structures shown by multicellular organisms}
\label{sec4}

\subsection{Multicellular organism}
Nature had invented multi-cellular systems, which are more favored for evolution in some case than the assembly of independent uni-cellular system [Process 5 of Figure \ref{fig1}]. During the long history of evolution, the lives which could have survived against the severe environment change must have equipped with a system which can maintain the structure and genome. Because the non-equilibrium thermodynamics of an open system does not necessarily guarantee the stability of the structure when the thermodynamic force X in Equation (\ref{eq:sigma2}) decreases by the environmental change, it is absolutely necessary for the living organism to prepare some protection mechanism of self-organized structure to survive. More accurately, a living organism had survived through the severe environment only when it happened to have a structure reversible for the environmental change. Because the reversibility is not achieved by any structures functioning in non-equilibrium state, the cell must have been protected by a system of equilibrium structure. This thought on crisis management is not surprising considering the fact that the cell membrane of a uni-cellular organization itself was originally not the product of non-equilibrium states. It has worked as the protecting wall of the self-replication machine of genes at early stage of life by some process shown in Figure \ref{fig6}, as an example.

\subsection{Morphology and entropy production of multicellular organisms}
Because each cell and its environment is not an independent open system far from the equilibrium any more, and the total entropy production is no more equal to the sum of that of each cell. The multi-cellular system is one thermodynamic system. The geometry of the collected cells can be in principle either one-dimensional array, two-dimensional sheets or three-dimensional aggregates. To obtain sufficient chemicals necessary for metabolic conditions, the collection of the cells needs enough surface. For this purpose, two-dimensional structure is best, because one dimensional array may be mechanically weak. Among various possible forms of two dimensional, the effective surface area of the collection of cells can be calculated for a sheet, a tube or a sphere of single layer of cells. The results tell us all the three forms of two-dimensional structure have 1/3 of the total surface area of individual cells.\footnote{Suppose that the surface area of a cell is not changed whether each cell is free or connected to each other. And, if the shape of a single cell becomes cubic for simplicity, when connected to each other in a plane, then 4 surfaces will be used for connection and only 2 surfaces out of 6 are free. The total free surface area of the cell assembly when connected in a plane is reduced to 1/3 of the total surface area of free cells. The situation is the same, even when the global structure of the plane is planar, tube-like or spherical.} However, the quantity of incoming chemicals into a small sphere of single cell by diffusion is severely reduced by the pinching effect of flow line, compared to that of incoming chemicals to the two-dimensional structure. Among them tube and sphere are superior to the plane because the they can form internal space, and tube is superior to the sphere because it is convenient for the liquid to flow from inlet to outlet. In fact, a tubular geometry is most often found in morphogenesis of primitive multi-cellular organism, such as coelenterates \cite{alberts83,lubarsky03}.

Reaction-diffusion chemical system is known to produce spatial pattern. This mechanism may give us a hint to understand an early stage of evolution to multi-cellular system, although genetic information and positional information play important roles in addition to this mechanism for the morphogenesis of present multi-cellular organism.

When the cells interact only by diffusion of the molecules, the entropy production may be written as,
\begin{equation}
  P = \frac{1}{T} \int [
    \sum_i A_i W_i + \sum_i \mu_i \sum_j \nabla (D_j \nabla \rho_{ij})] dV
\end{equation}
where $T$ is temperature, $A_i$ is the chemical affinity, $W_i$ is chemical reaction rate, and $\mu_i$ is the chemical affinity of $i$-th reaction. $D_i$ is the diffusion constant, and $\rho_{ij}$ is the density of $j$-th component of $i$-th reaction. By using the Brusselator \cite{prigogine68,glansdorff71} for the chemical reaction as a morphogenesis model of a multi-cellular system, It was found by simulation that the state of the chemical component of the maximum entropy production $P$ is most stable among others starting from various initial conditions \cite{shimizu83}. This observation will support a viewpoint that the [Process 5 of Figure \ref{fig1}] may be related the maximum entropy production principle, and that the pattern formation of multi-cellular system may be determined by the maximum entropy production principle, when the environmental condition such as chemical potential is maintained high enough.

Multi-cellular organism is characterized by differentiation. When a group of cells form a tubular structure, the cells at the openings experience different environmental situation from the cells far from openings. Among the various multi-cellular organization which mutation of genes would have created, tubular type of organism might have been selected.

\subsection{Altruistic behaviors based on switching in the organism from maximum entropy production principle to minimum free energy principle}
Some strategies obtained in the early history of evolution associated with the system such as germ cells in hydra or differentiation into spore-stalk in slime molds. Hydra shifts reversibly from vegetative growth to sexual reproduction by sacrificing amount of epithelial cells under the starving condition \cite{kaliszewicz11}. From life cycle of slime molds, we may also learn how multi-cellular organisms adapt in a restrictive environment \cite{rafols01}. When resources such as food are limited in the surrounding environment, population of single cells of independent amoebas is converted to a multi-cellular slugs, within which only the pre-spore cell become spore. During the phase of amoebas, differentiation to germline cells is not determined yet. In response to starvation, amoebas start to aggregates and only a certain ratio of cell in the aggregation become pre-spore cells \cite{rafols01}. This means that limited number of cells is potential for revival later in their life cycle and the others contributes only as structural components. In the higher multi-cellular organisms, germ cells are differentiated at earlier time of growth independently of the environmental condition.

Altruistic examples as seen in life cycle in slime molds are commonly observed in the multicellular society. Groups of cells sacrifice for the others to survive. Sacrifice of a fraction for the total does not appear to be consistent with maximum entropy production principle. However, the restrictive environment for living organisms forced the system to shift from a far-from equilibrium condition to an equilibrium condition. Thereby, thermodynamic principle for selection also switches from ‘the maximum entropy production principle’ to ‘the minimum free energy principle’. If the germ cells are more stable in the equilibrium state than the somatic cells, or if the germ cell can create not only germ cell itself but also somatic cells in the future mellow period, then one can be sure that the altruistically looking behavior of the multi-cell system follows the thermodynamics.

\subsection{Seeds and molecular structures of living state and dormant state}
Formation of seeds is one of the strategies of the differentiation to survive for plants. We want to know if the structure of the seeds is stable in the equilibrium state in which no chemicals penetrates through the shell. Surprisingly, some seeds of lotus, known as Oga Lotus \cite{ooga52}, which had been buried over 2000 years in an old ruin, successfully budded after immersed in water and finally bloomed. How do we thermodynamically understand this miraculous strength of life?

The most of multicellular animals become mobile for survival. However, particular animals adopt a strategy for survival similar to the seed of plants. The body of Tardigrades is the most famous for being able to survive in the extreme conditions that would be fatal to nearly all other known life forms, such as exposure to extreme temperatures and pressures (both high and low), air deprivation, radiation, dehydration and starvation \cite{kristensen11}. In the ‘open’ state the lotus seed or tardigrades body consist of structure which is metabolically active and undergoes events for living such as budding, growing, flowering, etc. Thermodynamically speaking, they are in an open system far from equilibrium and follows maximum entropy production principle in the normal living state. In the dormant or frozen state, on the other hand, a lotus seed is closed by a strong testa and exhibits no metabolic activity as if it were a stone. Thermodynamically it should be in a closed equilibrium system with the lowest Gibbs free energy, because the ‘seed’ is devoid of entry and exit of molecules and thus thermally separated from outside \cite{dragicevic11}.

\begin{figure}[H]
\centering
\includegraphics[width=12cm]{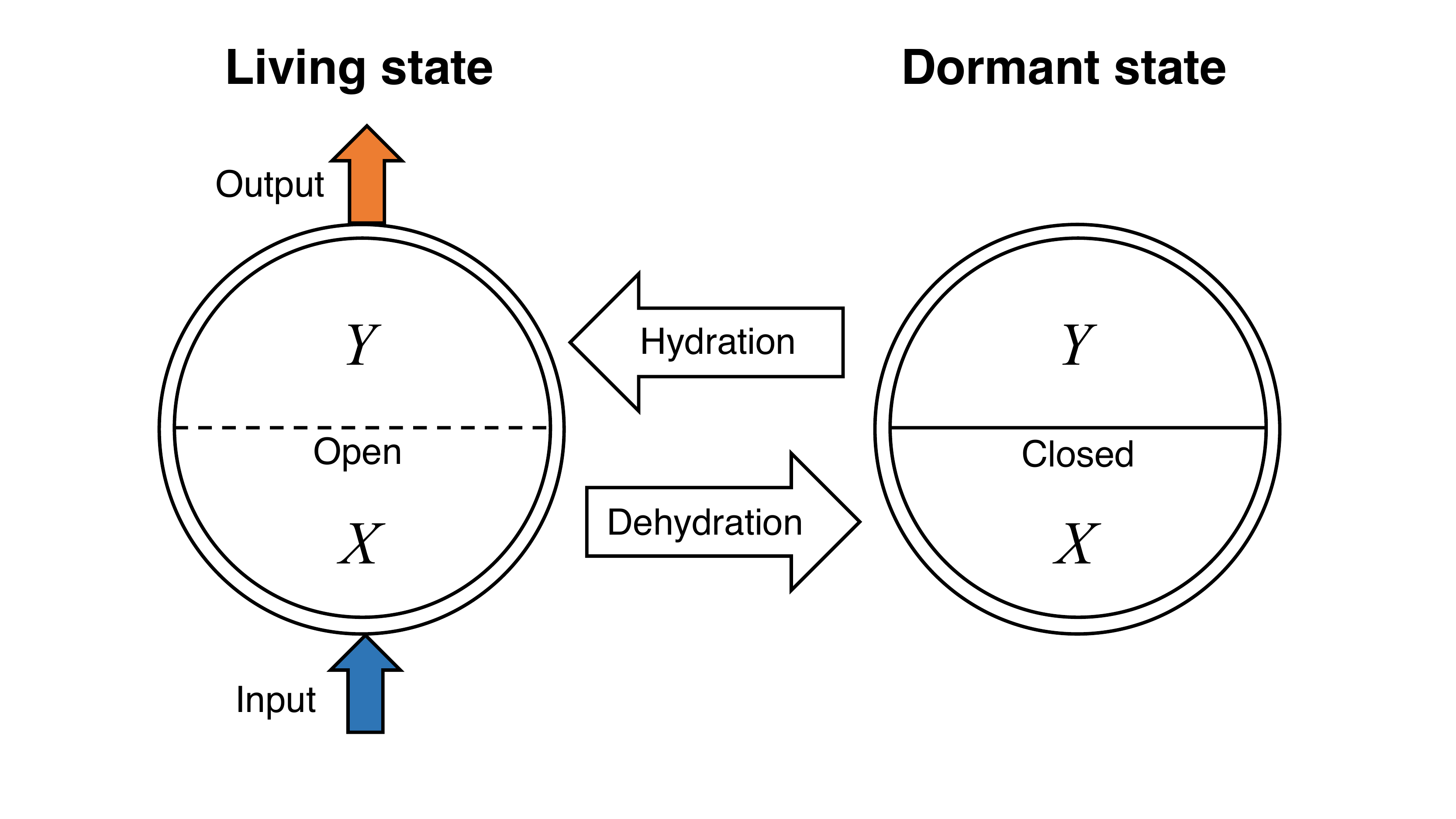}
\caption{
Switching between a living state and a dormant state of a cell exhibited by Oga lotus (\textit{Nelumbo nucifera Gaertner}) and Tardigrades, as examples. A self-organized living state is symbolically expressed by a chemical reaction $X$ to $Y$ separated in a chemical component space in an open non-equilibrium state with input and output.  By dehydration, the living system switches to a dormant state by closing the reaction wall between $X$ and $Y$.  The dormant state is stable for more than two thousand years for the Oga lotus seed. The process can be reversed by hydration to the living state. A possible mechanism may be related to the onset and offset of the activity of enzymes related for the reactions, although the physico-chemical mechanism has not yet clarified.
}
\label{fig7}
\end{figure}

Interestingly, being supplied with water, the seed quickly becomes ‘open’ status and start to enables pass of molecules though the shell. This event resembles an electric machine in which input of electricity changes the function without changing the macroscopic structure. The structure of dormant state of seeds, spores or a total body cannot be very different from when they are active in the ‘open’ state. At the same time the structure must be stable in a poor state as if it were a stone. In Figure \ref{fig7} is shown a schematic picture of switching between a living state and a dormant state of a cell, exhibited by Oga lotus and Tardigrades, as examples. Here, a self-organized living state is symbolically expressed by a chemical reaction $X$ to $Y$ separated in a chemical component space in an open non-equilibrium state with input and output.  By dehydration, the living system switches to a dormant state by closing the reaction wall between $X$ and $Y$.  The dormant state is stable for more than two thousand years for the Oga lotus seed. The process can be reversed by hydration to the living state. A possible mechanism may be related to the onset and offset of the activity of enzymes related for the reactions. Suggested glassy state \cite{buitink08} and disordered proteins \cite{boothby17}, which is very viscous in the absence of water and fluidic in the presence of water, may give the answer consistent with two different thermodynamic principles. Yet, no microscopic research of molecular structures of this transformation has been tried experimentally or theoretically to the authors’ best knowledge.

%%%%%%%%%%%%%%%%%%
\section{Discussion and Conclusion}
\label{sec5}

We presented in this paper a unified thermodynamic theory of the “Birth and Evolution of the life”. By inductive and deductive inference, we showed that the thermodynamic force, which has been known to increase the entropy production of the open system far from equilibrium, should have played a critical role for creating life and driving the evolution. In the early prebiotic stage, the entropy production is associated with chemical reactions of the molecular level, while in the successive stage after gene molecules are formed, the entropy production is due to the metabolic activity of each cell created by its own genetic information.

The generation of self-replication system and resulting exponential increase of number of organisms, which is certainly the most efficient pathway to increase entropy with no other comparison, is fundamentally distinct from non-life activities, and is reasonably called the birth of life. In fact, this is the basis of evolution of life to miraculous complexity of living phenomena. 

\begin{figure}[H]
\centering
\includegraphics[width=11cm]{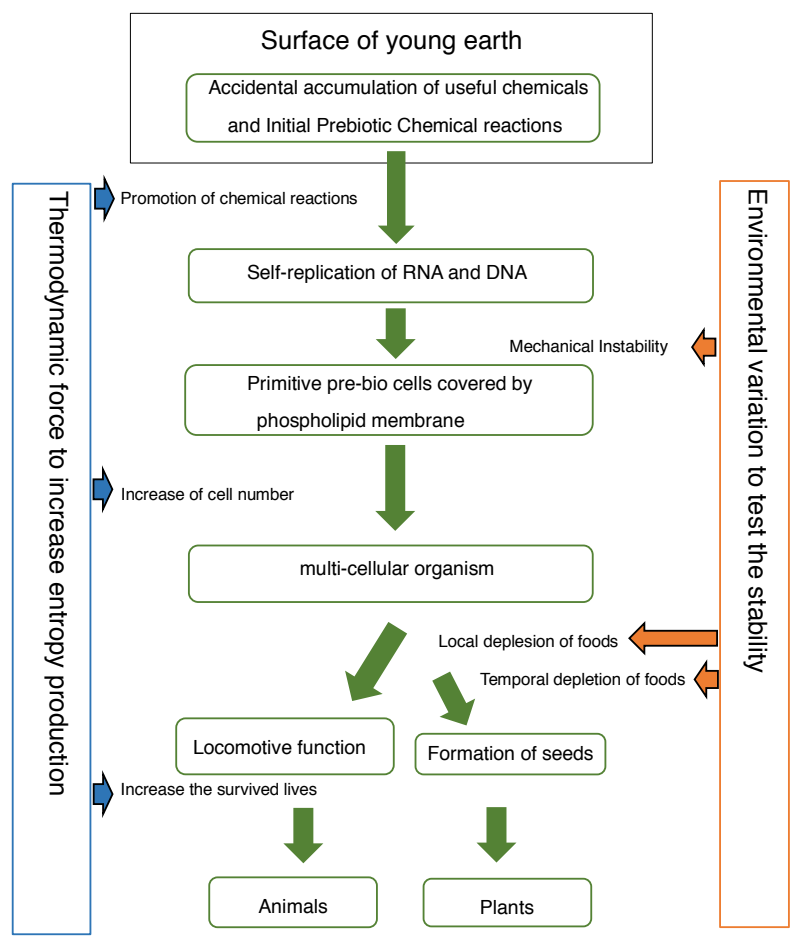}
\caption{
Thermodynamics for the birth, evolution and survival of life. Birth of life was achieved by the driving force of nonequilibrium thermodynamics, and the evolution was achieved by the competition between the thermodynamic force for increasing entropy production and destructive force of environmental change.
}
\label{fig8}
\end{figure}

The history of evolution, however, did not progress so simply. The fluctuation of environment sometimes did not keep the condition for the living organisms satisfying “far from equilibrium”. The foods for the organisms may have depleted temporarily or locally. Nature has also given the fluctuation among existing organisms by mutation. Some organisms had obtained seed-like cells or locomotive cells in the multi-cellular structure. These organisms which survived under starving condition, either by closing themselves or escaping to better environments, started growing their sizes or numbers again when the environment recover the “far from equilibrium” situation. Therefore, the driving force for the birth and evolution of life is achieved by the alternation of the non-equilibrium thermodynamic driving force for developing liveliness at a period of rich environment and protective structure of equilibrium thermodynamic force at a period of poor environment, as summarized in Figure \ref{fig8}.

%%%%%%%%5
%In short, life is a product of the thermodynamics which prevails in the universe together with the abundant useful chemical substance on the earth. The birth of our universe is in the present paradigm of cosmology described by “inflation”, in which some space-time region experienced exponential expansion. The universe, however, was not endowed with the self-replication mechanism like biological life.
%%%%%%%%

A question may arise; is this theory any better than a tautology that the statistically most probable phenomenon dominates?
%%%%%%%%5
%or that of “the survivals of the fittest” by Darwin?
%%%%%%%
Our answer is yes; we presented for the first time in this paper a thermodynamic driving force which initiated the birth and has driven the evolution, and thereby we could discuss the condition for the possibility of dominant processes in evolution. Birth of self-organized structures and competition among the various possible modes in nonlinear open systems far from equilibrium have been well established in various inanimate systems. We believe that it is important to explain the birth and evolution by the same principle which has been shown valid in the inanimate laboratory systems far from equilibrium.
%%%%%%%%%
%Darwin’s theory did not include origin of life nor driving force of the evolution.
%%%%%%%%

Secondly, is the directionality of the thermodynamic principle contradictory with the neutrality \cite{kimura71} of genetic mutation? Our answer is no. The mutation of gene is neutral. Among the various products by neutral mutation of genes, the organism of the maximum entropy production is selected, when the environmental condition is fertile.

Another question may be; is there predictability of the present theoretical work? We say yes in the relation of Process 6 of Figure \ref{fig1}, where the evolution of life finally created tools. Tools including fire has provided mankind the most efficient pathway for increasing entropy production, and have made mankind enjoy the results of using them. Only recently, they started worrying if the produced entropy might threaten themselves in future. A consistent thermodynamic viewpoint on life from Process 1 to 6, is potentially useful to predict the future of mankind.

Finally, if we admit a thermodynamic principle is the motive force for creation of life, shouldn’t there be other types of lives in some other stars which were made by the same principle but by different materials and evolved differently? Our answer is that there is no reason to say no.

%%%%%%%%%%%%%%%%%%%%%%%%%%%%%%%%%%%%%%%%%%
\vspace{6pt} 

%%%%%%%%%%%%%%%%%%%%%%%%%%%%%%%%%%%%%%%%%%
%% optional
%\supplementary{The following are available online at \linksupplementary{s1}, Figure S1: title, Table S1: title, Video S1: title.}


%%%%%%%%%%%%%%%%%%%%%%%%%%%%%%%%%%%%%%%%%%
\authorcontributions{
  %For research articles with several authors, a short paragraph specifying their individual contributions must be provided. The following statements should be used ``Conceptualization, X.X. and Y.Y.; methodology, X.X.; software, X.X.; validation, X.X., Y.Y. and Z.Z.; formal analysis, X.X.; investigation, X.X.; resources, X.X.; data curation, X.X.; writing--original draft preparation, X.X.; writing--review and editing, X.X.; visualization, X.X.; supervision, X.X.; project administration, X.X.; funding acquisition, Y.Y. All authors have read and agreed to the published version of the manuscript.'', please turn to the  \href{http://img.mdpi.org/data/contributor-role-instruction.pdf}{CRediT taxonomy} for the term explanation. Authorship must be limited to those who have contributed substantially to the work reported.
Conceptualization, Y.S.; investigation, Y.S., Y.D. and K.T.; visualization, Y.S. and K.T.; writing--Y.S., Y.D. and K.T.
}

%%%%%%%%%%%%%%%%%%%%%%%%%%%%%%%%%%%%%%%%%%
%\funding{Please add: ``This research received no external funding'' or ``This research was funded by NAME OF FUNDER grant number XXX.'' and  and ``The APC was funded by XXX''. Check carefully that the details given are accurate and use the standard spelling of funding agency names at \url{https://search.crossref.org/funding}, any errors may affect your future funding.}

%%%%%%%%%%%%%%%%%%%%%%%%%%%%%%%%%%%%%%%%%%
\acknowledgments{
One of the authors (Y.S.) acknowledges Prof. A. Libchaber, Prof. M. Obinata and Prof. M. Imai, and Dr. M. Robert for their useful discussions and kind introduction to the related references, and Dr. U. Tangen for his kind presentation of Eigen’s latest works.
}

%%%%%%%%%%%%%%%%%%%%%%%%%%%%%%%%%%%%%%%%%%
\conflictsofinterest{
  %Declare conflicts of interest or state ``The authors declare no conflict of interest.'' Authors must identify and declare any personal circumstances or interest that may be perceived as inappropriately influencing the representation or interpretation of reported research results. Any role of the funders in the design of the study; in the collection, analyses or interpretation of data; in the writing of the manuscript, or in the decision to publish the results must be declared in this section. If there is no role, please state ``The funders had no role in the design of the study; in the collection, analyses, or interpretation of data; in the writing of the manuscript, or in the decision to publish the results''.
The authors declare no conflicts of interest.
}

%%%%%%%%%%%%%%%%%%%%%%%%%%%%%%%%%%%%%%%%%%
%% optional
\appendixtitles{no} % Leave argument "no" if all appendix headings stay EMPTY (then no dot is printed after "Appendix A"). If the appendix sections contain a heading then change the argument to "yes".
\appendix
\section{Intuitive derivation of maximum entropy production principle}
\label{app1}
\unskip
According to the second law of the thermodynamics, the entropy of a closed system of particles with random distribution in momentum space always increases, and the ensemble approaches to Boltzmann distribution,
\begin{equation}
S(t+\Delta t) - S(t) > 0,
\end{equation}
because the larger the entropy is, the more probable the state is.

Suppose that this closed system is divided into a small closed subsystem and some reservoirs separated by the walls, each of which is in the equilibrium independently from each other. Furthermore, each reservoir is, by definition, very large and homogeneous defined by thermodynamic potentials $X$ such as temperature, pressure and chemical potentials, and will not produce entropy by itself. When the walls in contact with the small sub-system are replaced by semi-permeable walls at a time $t_1$, the small subsystem, which happens to have a thermodynamic potential different from the reservoirs in contact, will be driven to produce a thermodynamic current $J_i$ inside and thereby produces some entropy within the subsystem. Produced entropy may be extracted to the reservoirs to keep the subsystem steady, and the entropy of the total system will increase.

\begin{figure}[H]
\centering
\includegraphics[width=12cm]{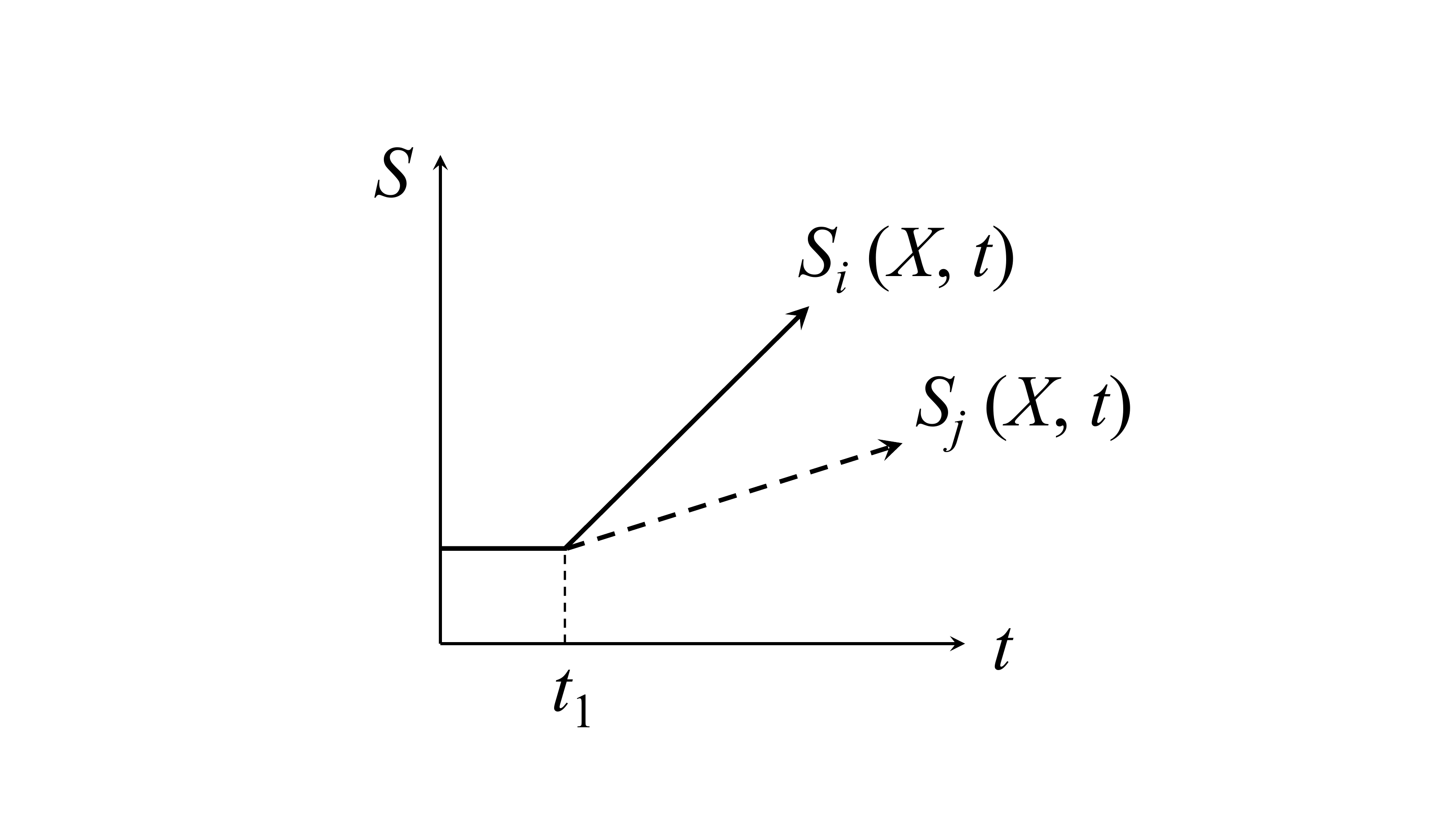}
\caption{
Thermodynamic mode selection by the maximum entropy production principle. The ordinate represents the total entropy including the subsystem and the reservoirs. A nonlinear subsystem far from equilibrium is generally known to have plural stationary modes. The figure schematically shows the total entropy change from $t = t_1$ at which the subsystem is opened to the reservoirs.  The mode corresponding to the highest entropy production $S_i(X,t)$ in the subsystem is selected thermodynamically among others.
}
\label{fig2}
\end{figure}

Let us discuss a subsystem hereafter, assuming for simplicity that all the subsystems are equal. Suppose a variety of active modes $i$ are possible for a given $X$ which is the thermodynamic potential far from equilibrium, the increase of the total entropy $S(t)$ for $t>t_1$ depends on the mode which the subsystem chooses. If $\sigma_i$, the entropy production per unit time of the $i$-th mode is greater than $\sigma_j$, that of the mode $j$, then $\Delta S(X,t) = \Delta t \cdot \sigma$ by definition we get,
\begin{equation}
  \Delta S_i(X,t) > \Delta S_j(X,t),
\label{eq2}
\end{equation}
and
\begin{equation}
S_i(X,t+\Delta t) > S_j(X,t+\Delta t) ~~~ \mathrm{for} ~~~ t > t_1.
\end{equation}
Then the mode $i$ of the local sub-system will be selected, because it is more probable. This is maximum entropy production principle which is a natural extension of the second law (see Figure \ref{fig2}).

Entropy production is equivalent to metabolic energy divided by the temperature for living organism. The former, however, is a concept which is useful quantity for living or nonliving systems, while the latter is a terminology used for living system only. That is why entropy production plays a fundamental law for the birth of life. Also, the difference is important for the [Process 6 of Figure~\ref{fig1}]. The energy dissipation for the body action to the environment using a tool is counted in metabolic energy, but the resultant increase of the environmental entropy is not counted in metabolic energy. However, it should also be counted as the entropy produced by the living object who used the tool.

\section{A simplest expression for chemical reactions realizing self-replication}
\label{app2}
\unskip
Among possible chemical reactions in the evolution, the RNA is considered to be the first pre-biotic self-replicating molecules which carry information \cite{gilbert86,szostak16}. Let us write RNA as $XX^*$, where $X$ is combined with $X^*$ which is a counter copy of $X$. The solution in a chemical reaction tank contains original RNA and the law nucleotides molecules $C$ and enzymes $E$, $E'$ and $E''$ which could be some proteins. 
The enzyme $E$ dissolves the original RNA $XX^*$ into $X$ and $X^*$. The reaction rate is $k_d$ with help of enzyme.
\begin{equation}
  XX^* \xrightarrow{k_d(E)} X + X^*
\label{eqA1}
\end{equation}
Subsequently, after the reaction (\ref{eqA1}) is over, $X$ and $C$ reacts to make a $XX^*$. The reaction rate is $k_c$ with help of $E’$.
\begin{equation}
  X + C \xrightarrow{k_c(E')} XX^*
\label{eqA2}
\end{equation}
Simultaneously with the reaction (\ref{eqA2}), $X^*$ and $C$ react to make a $X^* X$. The reaction rate should be the same that of (\ref{eqA2}),
\begin{equation}
  X^* + C \xrightarrow{k_c(E'')} X^*X
\label{eqA3}
\end{equation}
Since $XX^*$ is identical to $X^* X$, the value of $XX^*$ is doubled during the sum of the reaction times of (\ref{eqA1}) and (\ref{eqA2}) or (\ref{eqA1}) and (\ref{eqA3}), the number of RNA molecules, single stranded or double stranded increases exponentially with time constant $\tau = k_d^{-1}+(C k_c)^{-1}$.

The self-catalytic reaction rate of $XX^*(t)$ and therefore the entropy production increases exponentially. No other reactions can make the entropy production higher. Self-replication of RNA information by self-catalytic reaction was an indispensable step [Process 3 of Figure \ref{fig1}] for the evolution of life. The fastest growth of numbers of molecules \cite{eigen71,eigen13} can be achieved only by the self-replication ($m=1$) type among all possible chemical reaction systems of the type; $dN/dt = k N^m$. At the same time, this type of chemical reactions maximizes the entropy production, as we see in Section \ref{sec3}.

In the forgoing discussion, we did not discuss where the enzymes came from. If we include the reaction in which $X$ and $X'$ produce proteins which works as the enzymes, $E$ and $E'$, the network of reaction may be similar to one of the hyper-cycle \cite{eigen71,eigen13}.

%%%%%%%%%%%%%%%%%%%%%%%%%%%%%%%%%%%%%%%%%%
\reftitle{References}

\end{document}